\documentclass[%
 reprint,
 amsmath,amssymb,
 aps,
]{revtex4-2}

\usepackage{graphicx}
\usepackage{dcolumn}
\usepackage{bm}

\begin{document}

\preprint{APS/123-QED}

\title{New perspectives on future rip scenarios with holographic dark energy}

\author{Oem Trivedi$^{1}$ and Robert J. Scherrer$^{2}$}
\affiliation{$^{1}$International Centre for Space and Cosmology, Ahmedabad University, Ahmedabad 380009, India}
\affiliation{$^{2}$Department of Physics and Astronomy, Vanderbilt University, Nashville, TN, 37235, USA}

\email{Email: oem.t@ahduni.edu.in \\ Email: robert.scherrer@vanderbilt.edu }

\date{\today}

\begin{abstract}
We explore the asymptotic future evolution of holographic dark energy (HDE) models, in which the density of the dark energy is a function of a cutoff scale $L$.  We develop a general methodology to determine which models correspond to future big rip, little rip, and pseudo-rip (de Sitter) evolution, and we apply this methodology to a variety of well-studied HDE models.  None of these HDE models display little rip evolution, and we are able to show, under very general assumptions, that HDE models with a Granda-Oliveros cutoff almost never evolve toward a future little rip.  We extend these results to HDE models with nonstandard Friedman equations and show that a similar conclusion applies: little rip evolution is a very special case that is almost never realized in such models.
\end{abstract}

\maketitle

\section{Introduction}

The unexpected discovery of the Universe's late-time acceleration was a major surprise for the field of cosmology \cite{SupernovaSearchTeam:1998fmf}. Since then, extensive research has been conducted to explain this expansion phenomenon. The cosmological expansion problem has been approached from various angles, including conventional methods such as the cosmological constant \cite{Weinberg:1988cp,Lombriser:2019jia,Padmanabhan:2002ji}, as well as more unconventional theories such as modified gravity \cite{Capozziello:2011et,Nojiri:2010wj,Nojiri:2017ncd} and scenarios involving scalar fields driving late-time cosmic acceleration \cite{Zlatev:1998tr,Tsujikawa:2013fta,Faraoni:2000wk,Gasperini:2001pc,Capozziello:2003tk,Capozziello:2002rd,Odintsov:2023weg}. Furthermore, various approaches within the realm of quantum gravity have also contributed to addressing the cosmic-acceleration puzzle. These range from braneworld cosmology in string theory to theories like loop quantum cosmology and asymptotically safe cosmology \cite{Sahni:2002dx,Sami:2004xk,Tretyakov:2005en,Chen:2008ca,Fu:2008gh,Bonanno:2001hi,Bonanno:2001xi,Bentivegna:2003rr,Reuter:2005kb,Bonanno:2007wg,Weinberg:2009wa}. However, these efforts have highlighted certain discrepancies that suggest the limitations of our current understanding of the universe, with one of the most famous being the Hubble tension. This tension refers to disagreements in the values of the Hubble constant measured through detailed CMB maps, combined with baryon acoustic oscillations data and data from Supernovae Type Ia (SNeIa) \cite{Planck:2018vyg,riess2019large,riess2021comprehensive}. Therefore, the present epoch of the universe has presented us with a broad spectrum of questions and appears poised to become a domain where advanced gravitational physics will pave the way for a deeper comprehension of cosmology.

In the extensive array of proposed solutions for addressing the Dark Energy (DE) problem, one notable hypothesis is the holographic principle \cite{tHooft:1993dmi,Susskind:1994vu}, which holds significance in the context of quantum gravity. The core concept of the holographic principle suggests that the entropy of a system is not determined by its volume but rather by its surface area \cite{Bousso:1999xy}. Drawing inspiration from this idea, Cohen et al. \cite{Cohen:1998zx} proposed that in a quantum field theory, a short-distance cutoff is interconnected with a long-distance cutoff due to the limitation imposed by the formation of a black hole. In other words, if $\rho$ represents the quantum zero-point energy density caused by a short-distance cutoff, the total energy within a region of size $L$ should not exceed the mass of a black hole of the same size. Therefore, this leads to the inequality $L^3\rho\leq LM_{pl}^2$. The maximum allowable value for the infrared cutoff ($L_{_{IR}}$) is the one that precisely satisfies this inequality. Consequently, we have the relationship:
\begin{equation}
\label{simphde}
\rho=3 c^2L^{-2},
\end{equation}
where $c$ is an arbitrary parameter, and we are working in $m_{Pl} = 1 $ units.

This holographic concept has found wide application in cosmology, particularly in describing the late-time dark energy era, and is commonly referred to as holographic dark energy (HDE) (for an extensive review, see \cite{Wang:2016och}). From this perspective, the infrared cutoff, $L_{_{IR}}$, has its origins in cosmology and is the IR cut-off for a particular HDE model. Various other works have explored holographic dark energy from various aspects in recent years \cite{Nojiri:2017opc,Oliveros:2022biu,Granda:2008dk,Khurshudyan:2016gmb,Wang:2016och,
Khurshudyan:2016uql,Belkacemi:2011zk,
Zhang:2011zze,Setare:2010zy,Nozari:2009zk,Sheykhi:2009dz,
Xu:2009xi,Wei:2009au,Setare:2008hm,Saridakis:2007wx,Setare:2006yj,
Felegary:2016znh,Dheepika:2021fqv,Nojiri:2005pu, Nojiri:2017opc,Nojiri:2021iko,Nojiri:2020wmh}. In recent years, various other alternative forms of HDE have been proposed in recent decades. For example Tsallis HDE models are based on Tsallis' corrections to the standard Boltzmann-Gibbs entropy and is obtained after applying these corrections to  Black hole physics, resulting in the equation \begin{equation} \label{rtsa}
    \rho_{\Lambda} = 3 c^2 L^{-(4 - 2\sigma)},
\end{equation} where $\sigma$ is the Tsallis parameter, which is considered to be positive \cite{Tavayef:2018xwx} and we recover simple HDE in the limit $\sigma \to 1$.  On the other hand, Barrow's modification of the Bekenstein-Hawking formula led to the creation of Barrow HDE models and such models are described by the energy density \begin{equation} \label{rbar}
    \rho_{\Lambda} = 3 c^2 L^{\Delta - 2},
\end{equation} where $\Delta$ is the deformation parameter \cite{Saridakis:2020zol}, which can have a maximum value of $\Delta = 1 $ while in the limit of $\Delta \to 0$, one recovers the simple HDE. There are various other models too, like Renyi HDE \cite{moradpour2018thermodynamic} \begin{equation} \label{rren}
    \rho_{\Lambda} = \frac{3 d^2}{8 \pi L^2} (1 + \pi \delta L^2)^{-1},
\end{equation}
where $\delta$ is constant which can have positive or negative values, and for which the limit $\delta \to 0$ reduces it to the simple HDE scenario \eqref{simphde}. We also have the Kaniadakis HDE given by \cite{Drepanou:2021jiv}  \begin{equation} \label{rkan}
    \rho_{\Lambda} = 3 c^2 L^{-2} + 3 k^2 L^{2}
\end{equation} where k is a constant constrained as $-1<k<1$, and again for $k\to 0$ we recover the conventional HDE paradigm \eqref{simphde}. There are other HDE models besides these,  but \eqref{simphde}-\eqref{rkan} describe a wide variety of HDE models to investigate alongside the usual HDE model.

In recent times, a substantial body of literature has emerged focusing on the exploration of various types of singularities that may arise in the future
evolution of the Universe. The detection of late-time acceleration has significantly propelled such investigations \cite{Nojiri:2004ip,Nojiri:2005sr,Nojiri:2005sx,Bamba:2008ut,trivedi2022finite,trivedi2022type,odintsov2015singular,odintsov2016singular,oikonomou2015singular,nojiri2015singular,odintsov2022did,trivedi2024little,Trivedi:2023rln,Trivedi:2023wgg,trivedi2024recent,deHaro:2023lbq,lin2021effective,Odintsov:2022unp,Nojiri:2022xdo,Brevik:2021wzs}.  A particularly interesting class of such future events are rip scenarios, where the universe may proceed towards progressive disintegration in various capacities.
The central question addressed in this paper is the determination of
which of the various rip scenarios can take place in a universe with holographic dark energy and this is what we shall answer in this work. This question leads to the examination of a new way to approach rip scenarios, given the properties of HDE models. In section II we demonstrate a generalized Friedmann
equation that applies to these HDE models and derive conditions under which a future rip would occur in such models. In section III we apply those conditions to study big rip \cite{Caldwell:2003vq} and little rip \cite{Frampton:2011sp} evolution in a variety of HDE models. In section IV we examine HDE scenarios in non-GR cosmologies, such as the Chern-Simons and Randall-Sundrum type II braneworld.  Our conclusions are
summarized in section V. Our main conclusion is that little rip evolution is very difficult to achieve in any of these scenarios.

\section{Rip conditions with the HDE Friedmann equation}
Various future rip scenarios have been discussed in the literature.  Some of these scenarios take place in finite time while other require infinite time. One can summarize the various scenarios as follows \begin{itemize}
    \item Big rip (Type I singularity) : A well known scenario, where for $t \to t_{f}$, where $t_{f}$ is finite, we have both the effective energy density and pressure density of the universe diverging, $\rho_{eff} \to \infty $, $p_{eff} \to -\infty$,  while we also have a diverging Hubble parameter $H \to \infty$ \cite{Caldwell:2003vq}. This results in a scenario of universal destruction, where everything within the universe undergoes progressive disintegration. \cite{Caldwell:2003vq}.
    \item Little rip (LR) : Here the density, pressure and Hubble parameter become infinite as $t \to \infty$, \cite{Frampton:2011sp}. In this case all bound structures are eventually ripped apart but there is no finite time singularity. 
    \item Pseudo rip (PR) : In this case $H$ increases monotonically as $t \rightarrow \infty$, but it is bounded from
    above by the value $H_\infty$ so that  $H \rightarrow H_{\infty}$ as $t\to \infty$ \cite{Frampton:2011aa}.
    This is equivalent to asymptotic de Sitter expansion.
\end{itemize} 

Consider first the little rip. In Ref. \cite{Frampton:2011sp}, three types of parametrizations were examined and conditions were determined under which they led to a little rip. The three parametrizations involved the scale factor as a function of time, $a(t)$, pressure as a function of density $p(\rho)$ and density as a function of the scale factor $\rho(a)$. We will summarize those conditions briefly here.

As we are dealing with future scenarios, we will assume that the universe will eventually be dominated by dark energy so that the energy density and pressure will be given by $\rho_{DE} $ and $p_{DE}$, so from here onward we will drop the subscript.
We assume for now the standard Friedmann equation
\begin{equation} \label{Fre1}
    H^2 = \frac{\rho }{3},
\end{equation} 
with the continuity equation
\begin{equation} \label{Fre2}
    \dot{\rho} = -3 H ( \rho + p) 
\end{equation}
(These assumptions will be modified for the case of HDE models).

Consider first the case where $a$ is a specified function of $t$.
To avoid a big rip, we require $a(t)$ to be a nonsingular function for all $t$. Expressing $a$ as
\begin{equation}
\label{a(t)}
a = e^{g(t)},
\end{equation}
where $g(t)$ is nonsingular, the density is defined by equation (\ref{Fre1}) as $\rho = 3(\dot a/a)^2 = 3\dot g^2$. The condition for $\rho$ to increase with $a$ is $d \rho/da = (6/\dot a) \dot g \ddot g > 0$, satisfied if
\begin{equation}
\label{ddf}
\ddot g > 0.
\end{equation}
Therefore, all little-rip models follow equation (\ref{a(t)}), with a nonsingular $g$ satisfying (\ref{ddf}).

When the pressure is defined as a function of density, we consider an equation of state (EOS) given by
\begin{equation}
p = - \rho - f(\rho),
\end{equation} 
where $f(\rho) > 0$ ensures $\rho$ increases with the scale factor. To determine the existence of a future singularity, we integrate equation (\ref{Fre2}) to yield
\begin{equation}
a = a_0 \exp\left(\int \frac{d\rho}{3f(\rho)}\right),
\end{equation}
and equation (\ref{Fre1}) yields 
\begin{equation}
\label{trho}
t = \int \frac{d\rho}{\sqrt{3 \rho} f(\rho)}.
\end{equation}
For a big-rip singularity to occur, the integral in equation (\ref{trho}) must converge. If we take a power law for $f(\rho)$, namely
\begin{equation}
f(\rho) = A \rho^{\alpha},
\end{equation}
we find that a future singularity can be avoided for $\alpha \le 1/2$.

In the third scenario, $\rho$ is an increasing function of the scale factor $a$. We aim to find upper and lower bounds on the growth rate of $\rho(a)$ to determine if a big-rip singularity occurs. Defining $x \equiv \ln a$, we can express equation (\ref{Fre1}) as
\begin{equation}
t = \int \sqrt\frac{3}{\rho(x)} dx,
\end{equation}
and the condition to avoid a future big-rip singularity is
\begin{equation}
\label{condition}
\int_{x_0}^\infty \frac{1}{\sqrt{\rho(x)}} dx \rightarrow \infty.
\end{equation}

Specific models for the little rip are discussed then in \cite{Frampton:2011sp}.  Note, however, that general HDE models cannot be treated
with any of the parametrizations discussed in Ref. \cite{Frampton:2011sp} and presented here, because the standard Friedmann equation
does not apply to them.  This will lead to a different parameterization that can be examined in the context of little rip and big rip models.

Consider the choice for the IR cut-off scale $L$. An early suggestion was to consider a cutoff scale given by $L \to H^{-1}$. This choice aimed to alleviate the fine-tuning problem by introducing a natural length scale associated with the inverse of the Hubble parameter $H$. However, it was found that this particular scale resulted in an equation of state approaching zero, failing to contribute significantly to the current accelerated expansion of the universe.
An alternative approach involved utilizing the particle horizon as the length scale. This alternative resulted in an equation of state parameter higher than $-1/3$. However, despite this modification, the challenge of explaining the present acceleration remained unresolved.
Another option considered the future event horizon as the length scale.  Although the desired acceleration regime can be achieved in this case,
this approach raises problems with causality,
posing a significant obstacle to its viability.

To circumvent these difficulties, the Granda-Oliveros cutoff was proposed in Ref. \cite{Granda:2008dk}.  This cutoff is defined by the equation:
\begin{equation} \label{gocutoff}
L = (\alpha H^2 + \beta \dot{H})^{-1/2},
\end{equation}
where $\alpha$ and $\beta$ represent model parameters on the order of unity. The Granda-Oliveros cutoff successfully overcomes the issues encountered by previous proposals, providing a more robust framework for addressing the fine-tuning problem and explaining the accelerated expansion of the universe.

Combining this cutoff with Eq. (\ref{simphde}), we obtain a nonstandard form for the Friedman equation, namely
\begin{equation} \label{basic}
    H^2 = c^2 (\alpha H^2 + \beta \dot{H} ). 
\end{equation}
Clearly, the mathematical framework derived in Ref. \cite{Frampton:2011sp} cannot be straightforwardly applied here.

We can generalize this equation to
\begin{equation}
\label{generaleq}
    H^2 = f(H, \dot{H}).
\end{equation}
The simple HDE with the Granda-Oliveros cutoff is a special case of Eq. \eqref{generaleq}.
In fact, if one goes to non-GR cosmologies then in some cases the most general Friedmann equation can take the form \begin{equation} \label{mostgen}
    S(H) = f(H,\dot{H}).
\end{equation} 
We would like to here note that $S(H)$ here would be a function of the given background gravitational theory. For example, in the case of GR it would simply be $S(H) =H^2$ while for non-GR scenarios it can be much more involved as we shall see. We will return to Eq. (\ref{mostgen}) later; for now we will consider the evolution described by Eq. \eqref{generaleq}.
For most cases of interest, it will be possible to isolate the $\dot H$ term, allowing us to write
\begin{equation}
\dot H = g(H),
\end{equation}
where the function $g(H)$ is derived from Eq. (\ref{generaleq}). Note that the function g here has no connections with the function g in \eqref{a(t)}. Then we obtain
\begin{equation} \label{mainintegral}
    \int_{H_{i}}^{H_{f}} \frac{d H}{g(H)} = \int_{t_{i}}^{t_{f}} dt.
\end{equation}
Thus, for cosmologies which satisfy \eqref{generaleq} (or \ref{mostgen}) a new set of conditions determine whether a particular cosmological scenario allows for rip events. Those conditions are as follows \begin{itemize}
    \item Big rip: For the big rip we require that at some finite time $t_f$,
    we have $H_{f} \to \infty$ as $t \rightarrow t_f$. So for the big rip to occur in the context of Eqs. (\ref{generaleq}) or (\ref{mostgen}) we must have
    \begin{equation} \label{brcond}
    \int_{H_{i}}^{H_{f}} \frac{d H}{g(H)} \to \text{finite}, \quad \text{as} \quad H_{f} \to \infty.
\end{equation}
So a big rip occurs when the integral in \eqref{mainintegral} converges to a finite value as $H_{f} \to \infty$.
\item Little rip : For the little rip we require that $H_{f} \to \infty$ as $t \to \infty$.  Then for the little rip to occur in the context of (\ref{generaleq}) or (\ref{mostgen}) we must have 
\begin{equation}
	\label{lrcond}
    \int_{H_{i}}^{H_{f}} \frac{d H}{g(H)} \to \infty , \quad \text{as} \quad H_{f} \to \infty,
\end{equation} 
Hence for the little rip the integral in \eqref{mainintegral} diverges as $H_{f} \to \infty$.

\item Pseudo rip : For the Pseudo rip we have $H \to H_{f}$ as $t \to \infty$, where $H_{f}$ is some finite value. So for the pseudo rip to occur in the context of our generalized Friedmann equations we require 
\begin{equation}
\label{prcond}
\int_{H_{i}}^{H_{f}} \frac{d H}{g(H)} \to \infty, \quad \text{as} \quad H_{f} \to \text{finite value}.
\end{equation} 
Hence for the pseudo rip we need the integral in \eqref{mainintegral} to diverge for a finite $H_{f}$. 
\end{itemize}
From the point of view of HDE scenarios, conditions \eqref{brcond}-\eqref{prcond} provide a general view of the future rip status of any HDE paradigm with the Granda-Oliveros cutoff. It should also be noted that there are certain geodesic issues with the big rip singularity which may quite troublesome, but those have been discussed at length in other works \cite{Oikonomou:2018qsc,Harada:2021yul,Fernandez-Jambrina:2004yjt} and here we do not comment more on them. The question of whether or not this leads to preferring the other scenarios over the big rip is a bit open ended, especially when one can then make the case that the other rips are infinite-time scenarios and that might also be an issue. 
\\
\\
For the sake of completeness we shall also mention that there are other interesting rip possibilities like the Quasi rip \cite{Wei:2012ct} and Little sibling of the big rip \cite{,Bouhmadi-Lopez:2014cca}, but we do not consider them here. Besides the rip scenarios, we have other possibilities like the big brake singularity \cite{Gorini:2003wa,Kamenshchik:2007zj}. The big brake is special case of a sudden singularity \cite{Barrow:2004xh}, which is characterized by a diverging pressure density in a finite time, while the Big Freeze scenario \cite{bouhmadi2008worse} is characterized by diverging pressure and energy densities in finite time. These are again very interesting cases but are currently beyond the scope of our work and could make for very worthwhile future endeavours.
\section{Big rip and Little rip in HDE paradigms}

In this section, we will assume the Granda-Oliveros cutoff (Eq. \ref{gocutoff}) throughout and consider
the fate of various proposed HDE models.
We start by considering the simplest HDE scenario, given in \eqref{simphde},
which gives the Friedmann equation in the form
\begin{equation}
H^2 = c^2 ( \alpha H^2 + \beta \dot{H}).
\end{equation}
Isolating $\dot{H}$ in this scenario we see that \begin{equation} \label{simpintegral}
   \int_{H_{i}}^{H_{f}} \frac{ \beta c^2  dH}{H^2 ( 1- \alpha c^2)} = \frac{\beta c^2}{1-\alpha  c^2} \left(\frac{1}{H_{i}}-\frac{1}{H_{f}}\right) = \int_{t_{i}}^{t_{f}} dt.
\end{equation}
Clearly, $H_f \rightarrow \infty$ for a finite value of $t_f$, so this scenario
leads to a big rip for all cutoff parameters
$\alpha, \beta $ and $c$. Note that here we are not saying that if $t_{f}$ is finite then $H_{f}$ can never be finite, but what we are trying to emphasize is that $H_{f}$ diverges for a finite $t_{f}$ in the future, which signifies a big rip scenario. There are indeed interesting cases of singularities where like sudden and generalized sudden singularities \cite{Barrow:2004xh,Bamba:2008ut,Nojiri:2004pf} (the so-called type II and type IV singularities \cite{Nojiri:2005sx})  where divergences of these quantities have been studied in finite time as well, but here we are talking about the big rip ( a type I singularity). In contrast to the sudden singularities, there is complete divergence of the pressure and energy densities, alongside the scale factor in the future of the universe and here we are reaching towards that sort of a scenario as in a finite time, with a diverging $H_{f}$ one would reach a DE energy density $\rho_{DE} \to \alpha H_{f}^2 + \beta \dot{H_{f}} \to \infty$, which would consequently mean that the pressure density also diverges ( $p=w \rho$). One can indeed study other finite time singularities like these, but as we are interested in rip scenarios we are not pursuing it here.
\\
\\
Now consider the Tsallis model for which
\begin{equation}
f(H,\dot H) = c^2 (\alpha H^2 + \beta \dot{H})^{(2- \sigma)},
\end{equation}
from which we get
\begin{equation} \label{tsaintegral}
    \int_{H_{i}}^{H_{f}} \frac{\beta dH}{\left[ \left(H/c \right)^{2/(2-\sigma)} - \alpha H^2 \right]} = \int_{t_{i}}^{t_{f}} dt
\end{equation}
The Barrow model (Eq. \ref{rbar}) yields a similar
expression for $f(H, \dot H)$, corresponding to
\begin{equation}
f(H, \dot{H}) = c^2 (\alpha H^2 + \beta \dot{H})^{(2- \Delta)/2},
\end{equation}
from which we have
\begin{equation} \label{Barintegral}
\int_{H_{i}}^{H_{f}} \frac{\beta dH}{\left[ \left( \frac{H}{c} \right)^{4/(2-\Delta)} - \alpha H^2 \right]} = \int_{t_{i}}^{t_{f}} dt.
\end{equation}

Note that both the Tsallis and Barrow models correspond to a general evolution of the form
\begin{equation}
\label{gendotH}
\dot H = (1/\beta)\left[\left({H}/{c} \right)^{n} - \alpha H^2 \right],
\end{equation}
so
\begin{equation} \label{genintegral}
\int_{H_{i}}^{H_{f}} \frac{\beta dH}{\left[ \left({H}/{c} \right)^{n} - \alpha H^2 \right]} = \int_{t_{i}}^{t_{f}} dt,
\end{equation}
although the allowed values for $n$ will be quite different for these two models.  We will examine the general evolution
given by Eq. (\ref{genintegral}) and then work backwards to apply it to the Tsallis and Barrow models.

While the integral in Eq. (\ref{genintegral}) can be derived exactly in terms of hypergeometric functions, this result does not yield much insight.  Instead,
note first that Eq. (\ref{genintegral}) allows for two different sets of solutions, depending on the sign of $\dot H$. 
Solutions with $\dot H < 0$ in Eq. (\ref{gendotH}) cannot evolve toward a future rip singularity, so we will consider only the case where $\dot H > 0$.  Then in Eq. (\ref{genintegral}), the integral
will converge whenever $n \ge 2$, indicating a big rip singularity at a finite time.  For $n < 2$, 
$H$ goes to a finite value as $t \rightarrow \infty$, corresponding to a pseudo-rip.

Now we can compare these results to the Tsallis and Barrow models.  For the Tsallis model, the big
rip condition, $n \ge 2$ corresponds to $1 \le \sigma < 2$.  Thus, a future big rip is possible in
the Tsallis HDE model for this parameter range.  In the Barrow model, $\Delta$ is restricted to lie
in the range $0 < \Delta \le 1$, which corresponds to $2 < n \le 4$.  Thus, the Barrow model
can also lead to a future big rip singularity.

The Renyi model \eqref{rren} is the case in which $f(H, \dot{H})$ takes the form \begin{equation}
    f(H,\dot{H}) = \frac{d^2}{8 \pi} (\alpha H^2 + \beta \dot{H} ) \left[ 1 + \frac{\pi \delta}{(\alpha H^2 + \beta \dot{H} )} \right]^{-1},
\end{equation}
from which we can get the integral
\begin{equation} \label{renintegral}
    \int_{H_{i}}^{H_{f}} \frac{\beta dH}{H^2 \left[ \left( \frac{4 \pi}{d^2} - \alpha + \frac{4 \pi}{d^2} \sqrt{1 + \frac{d^2 \delta}{2 H^2}} \right) \right]} = \int_{t_{i}}^{t_{f}} dt.
\end{equation}
This integral can be expressed analytically, but it is easy to see that 
as long as $\dot H > 0$, the integrand on the LHS will scale as $dH/H^2$ as $H \rightarrow \infty$,
giving a convergent integral and a future big rip, independent of the model parameters.

Finally consider the Kaniadakis model (\ref{rkan}).
The Kaniadakis model \eqref{rkan} is the case in which $f(H, \dot{H})$ takes the form \begin{equation}
    f(H,\dot{H}) = c^2 (\alpha H^2 + \beta \dot{H} ) + \frac{k^2}{(\alpha H^2 + \beta \dot{H} )}
\end{equation} from which we derive \begin{equation}
    \int_{H_{i}}^{H_{f}} \frac{2 \beta c^2}{H^2 \left[1-2\alpha c^2 + \sqrt{1 - \frac{4k^2 c^2}{H^4}} \right]} dH = \int_{t_{i}}^{t_{f}} dt
\end{equation} 
We see that for the case in which $\dot H > 0$, the integral over the Hubble parameter always converges, leading to a future big rip.

The HDE models we have examined here generically evolve toward a future big rip, and in no case do we see evolution toward a little rip.  We now demonstrate a general result:  HDE models with the Granda-Oliveros cutoff
do not evolve toward a little rip except for very special choices of the functional form for $\rho_\Lambda(L)$.
Consider a general form for $\rho_\Lambda$ as a function of $L$:
\begin{equation}
\rho_\Lambda = f(L),
\end{equation}
with $L$ given by Eq. (\ref{gocutoff}).  We have examined various proposals for $f(L)$; now we allow it to be a free function.  Then we have $H^2 = f(L)$, and using the Grand-Oliveros cutoff we get
\begin{equation}
f^{-1}(H^2) = (\alpha H^2 + \beta \dot H)^{-1/2},
\end{equation}
where $f^{-1}$ is the inverse function corresponding to $f(L)$.  Our integral relation between $H$ and $t$
becomes
\begin{equation}
\label{generalevolution}
    \int_{H_{i}}^{H_{f}} \frac{\beta dH}{[f^{-1}(H^2)]^{-2} - \alpha H^2} = \int_{t_{i}}^{t_{f}} dt.
\end{equation}
Now consider the asymptotic behavior of the integrand in the limit of large $H$.
If $[f^{-1}(H^2)]^{-2}$ increases faster than $H^2$ as $H \rightarrow \infty$, the integral will converge, corresponding to a big rip.
If $[f^{-1}(H^2)]^{-2}$ increases more slowly than $H^2$, then the denominator in Eq. (\ref{generalevolution})
will go to zero at some finite value of $H$, so that $H$ will approach a constant as $t \rightarrow \infty$.
This corresponds to a pseudo-rip, or equivalently, asymptotic de Sitter evolution.

The one possibility
of a little rip corresponds to the case where $[f^{-1}(H^2)]^{-2}$ evolves as $\alpha H^2 + g(H)$ as $H \rightarrow
\infty$, where $\int dH/g(H)$ diverges.  However, such behavior is rather contrived and does not correspond
to any models proposed thus far in the literature.  As an example, the simplest case producing a little
rip is $[f^{-1}(H^2)]^{-2} = \alpha H^2 + kH$, where $k$ is a constant.  Working backwards, this corresponds
to
\begin{equation} \label{cutform}
\rho_\Lambda = \frac{3}{\alpha}\left(\frac{1}{L^2} - k\right).
\end{equation}
Thus, the little rip is not a possible future evolution except for a very small class of Granda-Oliveros
HDE models. We note that the cutoff form needed in \eqref{cutform} is within the realm of the generalized Nojiri-Odintsov cutoff \cite{Nojiri:2005pu,Nojiri:2017opc,nojiri2019holographic}, where the HDE energy density in general could have the form \begin{equation} \label{nocutoff}
    \rho = 3 c^2 L(H,\dot{H},\ddot{H},...L_{p},L_{f}, \dot{L_{p}},\dot{L_{f}}...) 
\end{equation} where $L_{p}$ and $L_{f}$ refer to the particle horizon and future event horizon \begin{equation} \label{lplf}
    L_{p} = a \int_{0}^{t} \frac{dt}{a} \quad , \quad L_{f} = a \int_{a}^{\infty} \frac{dt}{a}
\end{equation} In general one can have HDE models where the density could be of the form \eqref{nocutoff} being general functions of H, particle horizon and event horizon scales \eqref{lplf}, so it would not be completely lost on one to find various possibilities of the generalized cutoff which could accommodate ( or remove) multiple rip scenarios in the usual HDE model \eqref{simphde} or its extensions like Tsallis etc. But one in general struggles to motivate a particular form for the function in \eqref{nocutoff}, even if one wants to allow for the generality of the cutoff.  
\\
\section{Generalizing to non-GR cosmologies}

So far we have considered the HDE models in the framework of the standard Friedman equation, corresponding
to Eq. (\ref{generaleq}).  Here we will extend our argument to several proposed models in which
the Friedman equation is modified, resulting in an expression of the form given in Eq. (\ref{mostgen}).

As a first example, we consider the Chern-Simons cosmology, which corresponds to \cite{gomez2011standard}
\begin{equation} \label{chernfried}
    S(H) = H^2 - \gamma H^4
\end{equation}
where $\gamma $ is a free parameter for the model, which takes positive values.
In the context of the
simple HDE model (Eq. \ref{simphde}) with the Granda-Oliveros cutoff, we get the integral \begin{equation}
    \int_{H_{i}}^{H_{f}} \frac{\beta c^2 dH}{H^2\left( 1 - \gamma H^2 - c^2 \alpha \right)} = \int_{t_{i}}^{t_{f}} dt
\end{equation}

Again, this can be integrated exactly, but we are interested in the asymptotic behavior of the integral.
We see that the integral diverges for a finite value of $H_f$.  Thus, these models evolve generically to
a pseudo-rip (asymptotic de Sitter expansion).

Another important model with a nonstandard Friedman equation is the RS-II braneworld
which corresponds to \cite{Randall:1999vf,Deffayet:2000uy,Gonzalez:2008wa,Escobar:2011cz}
\begin{equation}
    f(H,\dot{H}) = c^2 (\alpha H^2+ \beta \dot{H}) \left( 1 + \frac{3 c^2}{2 \lambda} (\alpha H^2+ \beta \dot{H}) \right),
\end{equation}
which gives
\begin{equation}
    \int_{H_{i}}^{H_{f}} \frac{3 c^2 \beta dH}{\lambda \left( \sqrt{\frac{6 H^2}{\lambda} +1} - 1 \right) - 3 c^2\alpha H^2} = \int dt
\end{equation}
Again, we see that the integral on the left-hand side diverges at finite $H$, so that these models also evolve
to asymptotic de Sitter expansion.

While there are essentially an infinite number of nonstandard Friedman equations that one might conceivably incorporate into
the HDE framework, we can show, rather surprisingly, that we expect little rip evolution to be rather rare in all of these cases
as well.  Consider a generic evolution of the form
\begin{equation}
g(H^2) = \rho_\Lambda = f(L)
\end{equation}
where $L$ is taken to given by the Granda-Oliveros cutoff.  Then we can write
\begin{equation}
f^{-1}(g(H^2)) = (\alpha H^2 + \beta \dot H)^{-1/2}
\end{equation}
and we get
\begin{equation}
\label{generalevolution2}
    \int_{H_{i}}^{H_{f}} \frac{\beta dH}{[f^{-1}(g(H^2))]^{-2} - \alpha H^2} = \int_{t_{i}}^{t_{f}} dt.
\end{equation}
Then our argument carries over from the previous section exactly as it did to Eq. (\ref{generalevolution}).
Except for very special choices of $f$ and $g$, asymptotic little rip evolution is impossible.
We also note that if one were to consider $\alpha = 2 \beta $, then one recovers the Ricci Holographic dark energy model \cite{Zhang:2009un}, and the conclusions that we have reached at remain unchanged even in this HDE paradigm which further highlights the generality of these results.
\section{Conclusions}
In this work we have considered a very general approach to future rip scenarios with holographic dark energy.  Using this
new methodology, we have shown how to determine, in a simple way, whether specific HDE models correspond to little rip,
big rip, or future asymptotic de Sitter (pseudo-rip) evolution.  All of the specific models we have examined correspond
to a future big rip or de Sitter evolution, with none of them yielding a little rip.

More generally, in HDE models with a Granda-Oliveros cutoff, it is possible to show that little rip evolution is a very
special case, requiring a rather contrived choice for $\rho_\Lambda$ as a function of the cutoff $L$.  This result
arises mathematically because $L$ for the Granda-Oliveros cutoff is a sum of $\dot H$ and $H^2$, so that
$\dot H$ always enters into the evolution equations in the form of this sum.

Rather surprisingly, this result carries over to HDE models with nonstandard Friedman equations, such as the Chern-Simons
or Randall-Sundrum type II braneworld cosmologies.  Even when one allows for an arbitrary relationship between $H^2$
and $\rho_\Lambda$, the specific form for $L$ with the Granda-Oliveros cutoff renders it difficult to construct models
corresponding to a little rip.

While we have considered primarily HDE models with the Granda-Oliveros cutoff, it would be interesting
to investigate how general our results are.  It may be that the little rip is a very special case,
requiring particular fine-tuning in any cosmological model.  This question is worthy of further exploration.

\section*{Acknowledgements}
The authors would like to thank Alexander Oliveros for various helpful discussions on holographic dark energy. We would also like to thank the anonymous referee for their insightful comments on our work.
\bibliography{apssamp}

\end{document}